\documentclass[rnote]{aa} 
\usepackage{graphicx}
\usepackage{txfonts}
\begin{document}

\title{High spatial resolution imaging of the star with a
transiting planet WASP-33}


   \author{A. Moya
          \inst{1}
          \and
          H. Bouy
          \inst{1}
          \and
          F. Marchis
          \inst{2}
          \and
          B. Vicente
          \inst{3}
          \and
          D. Barrado
          \inst{4,1}
          }

   \institute{Departamento de Astrof\'{\i}sica, Centro de
     Astrobiolog\'ia (INTA-CSIC), PO BOX 78, 28691 Villanueva de la
     Ca\~nada, Madrid, Spain\\ \email{amoya@cab.inta-csic.es}
     \and Department of Astronomy at UC-Berkeley, 601 Campbell Hall, Berkeley 
          CA 94720 USA Carl Sagan Center at the SETI Institute, 189 
          Bernado Av, Mountain View CA 94043, USA
     \and Instituto de Astrof\'isica de Andaluc\'ia (CSIC), Glorieta de
          la Astronom\'ia S/N, 18008, Granada, Spain
     \and German-Spanish Astronomical Center, Observatorio de Calar Alto, 
          C/ Jes\'us Durban Remon 2-2, 04004 Almeria, Spain}

   \date{}

 
  \abstract
   {The recent discovery of a transiting planet around WASP-33,
     the misalignment of the stellar rotation axis and the planet orbit,
     the possible existence of an additional planet in the system, and
     the presence of $\delta$ Scuti pulsations in the hosting star
     make this system a highly interesting object to help us understand the
     origin and evolution of giant planets orbiting very close a star.}
   {High spatial resolution imaging
   with an adaptive optics system on the W.M. 10m-Keck II telescope allows the 
   study of the presence of the predicted additional planet,
   and any other objects, constraining the possible formation scenarios
   of the system.}
   {In November 2010, we recorded high spatial resolution
   images from 1 to 2.5 $\mu$m using the W.M. 10m-Keck II telescope and its
   adaptive optics system, obtaining broad (Jc,Hc, and Kc) and narrow
   band (FeII) images of the system. After data reduction, the
   contrast and angular resolution
   provided by this instrument allowed us to
   constraint the multiplicity of this system and to detect one
   potential companion.}
   {We have found a new object at a distance of
   1.961$"$ $\pm$ 0.003$"$ from the WASP-33, with a position 
   angle of 276.32$\pm$0.24 deg. It could be a dwarf star/ brown dwarf 
   or an extragalactic object. In the first case, and assuming the same 
   distance from the Sun, the object is 227~AU from the
   central star. On the other hand, no additional objects have been
   found. This constraints the possible objects in the system,
   depending on its age and masses.}
   {A potential companion object to WASP-33 was obtained. The
   gravitational link must be confirmed, but this object could influence
   the evolution of the planetary system depending on its orbital 
   eccentricity. We have ruled-out the
   existence of additional objects. For example, objects of 0.8 $M_\odot$ 
   at projected physical distances greater than [2, 5] AU, 0.3 
   $M_\odot$ at projected physical distances greater than [11,
     18] AU and 0.072 $M_\odot$ at projected physical distances 
   greater than [18, 75] AU, depending on the age of the system.}

   \keywords{stars: individual (WASP-33, HD~15082) - stars: binaries -
   techniques: high angular resolution}

   \maketitle
%

\section{Introduction}

The discovery of hundreds of exoplanets in the last few years, as a
result of the increasing number of observational techniques, and their
accuracies, has opened a number of questions about the origin and
evolution of planetary systems. The number of theories developed to
explain the observational facts has also increased in a similar
way. Due to observational biases, a large number of massive gas
planets orbiting close to the hosting star have been reported. That
stimulated the development of several theories explaining the assumed
planet migration, including disk-planet interaction models (Ida \&
Lin, \cite{Idalin}), planet-planet scattering models (Nagasawa et al.,
\cite{Nagasawa}, Chatterjee et al., \cite{Chatterjee}), or Kozai
  mechanism plus tidal dissipation models (Wu, Murray \& Ramsahai,
\cite{Wu}). The alignment of the planet's orbit and the stellar
rotation axis, measured mainly using the Rossiter-McLaughlin effect,
is a key disentangling, and discriminating between the different
formation and migration scenarios. Thanks to the most recent
observational facts, the misalignment found in some systems can be
explained by the presence of other massive bodies (Narita et al.,
\cite{Narita}), or can be related to the stellar effective
temperature, and therefore with its internal structure (Winn et al.,
\cite{Winn}). This last prediction suggests a link between the
planet-star misalignment and the absence of an outer highly developed
convective zone, that is, it is restricted to stars with
$T_{eff}>6200\,^{\b{o}}K$, approximately. The study of hot stars with
planets is imperative to the understanding of these scenarios.

The discovery of exoplanets around A stars using Radial-Velocities
(RV) is not an easy task. The mass of these stars only allows the
detection of massive planets orbiting close to the star. WASP-33
(HD15082, $V=8.33$, $v\sin i=86$ km s$^{-1}$, Christian et al.,
\cite{chris}) is the only $A$ star known with a transiting
planet. This planet has been confirmed using RV (Collier-Cameron et
al., \cite{wasp33}); that means that the planet is massive. In this
latter work, the authors: 1) refine the stellar and planetary physical
parameters, 2) confirm the orbital period of the planet (1.22 d) and,
3) report a misalignment between the stellar and orbital spin axis. In
addition, they suggest possible stellar $\delta$ Scuti pulsations,
something not surprising since the star is located in the HR $\delta$
Scuti Instability Strip (Grigahc\`ene et al. \cite{ahmed}). These
pulsations where confirmed by Herrero et al. (\cite{puls}), and they
can help in understanding the internal structure of the star. Herrero et
al. (\cite{puls}) found a coincidence between the pulsating period and
a multiple of the orbital period. All these characteristics make
WASP-33 a very interesting object from the point of view of both giant
planets formation and evolution and stellar structure and evolution.

\begin{table}
\begin{tabular}{ccccc}
\hline
Object & Band & Value (mag) & Error (mag)& Source \\
\hline
 & J & 7.581 & 0.021 & \\
WASP-33 & H & 7.516 & 0.024 & 2MASS\\
 & Ks & 7.468 & 0.024 & \\
\hline
& $\Delta$Jc & 6.62 & 0.03 & \\
 & $\Delta$Hc & 6.13 & 0.02 & \\
New & $\Delta$Kc & 5.69 & 0.01 & This work\\
(candidate) & $\Delta$Jc-$\Delta$Kc & 1.04 & 0.04 & \\
 & $\Delta$Jc-$\Delta$Hc & 0.55 & 0.05 & \\
 & $\Delta$Hc-$\Delta$Kc & 0.49 & 0.03 & \\
\hline
\end{tabular}
\caption{Photometric colours observed for WASP-33 and the difference
  between the object discovered and WASP-33 in the filters Jc, Hc and
  Kc.}
\label{char}
\end{table}

The presence of additional giant planets orbiting the hosting star
would provide a better understanding of the planet formation
scheme. In Collier-Cameron et al. (\cite{wasp33}), the possible
existence of an additional object is pointed out taking into account
the RV residuals after the planet and pulsations analysis. In the
present study, we have searched for additional WASP-33's companion
objects through direct imaging.

\section{Observations and data processing}

We observed WASP-33 using the Adaptive Optics system of the Keck-II
telescope and its NIRC2 infrared camera ($1024\times 1024$
pixels). With a V-magnitude of 8.3, the star is bright enough to be
used as a reference of the wavefront analysis. The data collected on
November 29 2010 were recorded in Jc (bandwidth of 0.020 $\mu$m
centred on 1.213 $\mu$m), Hc (bandwidth of 0.023 $\mu$m centred on
1.580 $\mu$m) and Kc (bandwidth of 0.030 $\mu$m centred on 2.271
$\mu$m), under good seeing conditions but high humidity. For each
filter, we recorded 3 images at different positions of the
detector. Each image is composed of 5 in Jc, 5 in Hc and 20 in Kc,
co-added frames with an individual exposure time of 2.9s. The angular
resolution in Kc-band is estimated in 55 milli-arcsec (mas), very
close to the diffraction limit of the telescope (47 mas).  Additional
observations were recorded on November 30 in Kc and FeII (bandwidth of
0.026 $\mu$m centred on 1.645 $\mu$m) filters. The data set is
identical with 3 images composed of 6 in Kc and 20 in FeII co-added
frames with an individual exposure time of 3s.  Since the atmospheric
conditions were not optimal (fog and high altitude cirrus) the angular
resolution in Kc band is estimated in 100 mas. For both nights, the
observations were performed at $\sim$10 UT with an optimal low
air-mass of 1.1. We estimated the seeing at the beginning of the first
night at $\sim$0.7$''$ in visible. The Strehl ratio was estimated at
$\sim$40\% on the first night using a Kc simulated PSF.

\begin{figure*}
\includegraphics[width=15cm]{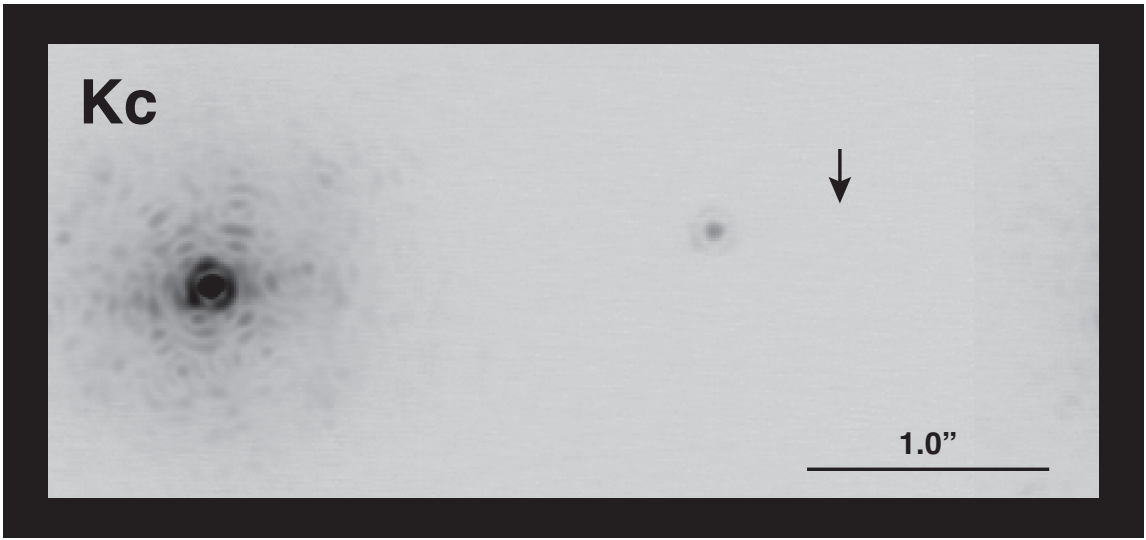}
\caption{Direct imaging of WASP-33 in the Kc-band. The new object,
  located at $\sim$2 arcseg is marked with an arrow.}
\label{imagen}
\end{figure*}

The data were processed using standard procedures with the Eclipse
software package (Devillard, \cite{Devillard}). The individual images
were flat-field corrected and then registered and median
combined. Aperture photometry of the 2 sources present in the field
was extracted using the IRAF \footnote{IRAF is distributed by the
  National Optical Astronomy Observatory, which is operated by the
  Association of Universities for Research in Astronomy (AURA) under
  cooperative agreement with the National Science Foundation.} daophot
package. An aperture of 20 pixels and a sky annulus between 20-25
pixels were used. The flux ratio between the bright primary and the
wide companion was computed, and the absolute magnitude of the wide
companion was derived in each of the broad band filters using the
2MASS magnitudes of the primary (Skrutskie et al. \cite{2MASS}).

No astrometric standard field was obtained during the night, and the
NIRC2 pixel scale and orientation cannot be accurately
calibrated. Ghez et al. (\cite{ghez}) found an average platescale of
9.963$\pm$0.006 mas/pixel over the period 2004 and 2008. We have
adopted this value for our measurements. The uncertainty on the
position angle of the camera on the sky is also unknown. It depends
mostly on eventual engineering adjustments which can introduce slight
rotations in the camera. Ghez et al. (\cite{ghez}) found systematic
offsets of up to 0.2 deg with respect to the rotation keywords given
by the control system. This offset has been also added to the
measurement uncertainties. No distortion correction was applied to our
measurements.

\section{Result 1: A new companion candidate to WASP-33}

In Fig. \ref{imagen}, the final image after basic data-processing is
shown in the Kc-band. Here we can see the presence of an object
1.961$"$ $\pm$ 0.003$"$ distant from the star and at a position angle
of 276.32 $\pm$ 0.24 deg. The uncertainty was computed from the
1$\sigma$ dispersion of the individual measurements in each filter
plus the above mentioned instrumental uncertainties (platescale and
position angle of the camera on the sky). No other object was found by
visual inspection. This is an unknown object. The first part of this
study is devoted to the analysis of this object and its possible
membership to the WASP-33 system. Table 1 shows the photometric
colours observed, including those of WASP-33 as reference. No
photometric standard star was obtained during the night. In order to
assess the nature of the companion we hereafter assume that the
flux-ratio in the Jc, Hc and Kc Mauna Kea filters can be applied in
the corresponding 2MASS filters system. Therefore, in the absence of
confusing issues, we use the 2MASS filters notation to label our
results when WASP-33 and the new candidate are compared.

\begin{figure}
\includegraphics[width=8.5cm]{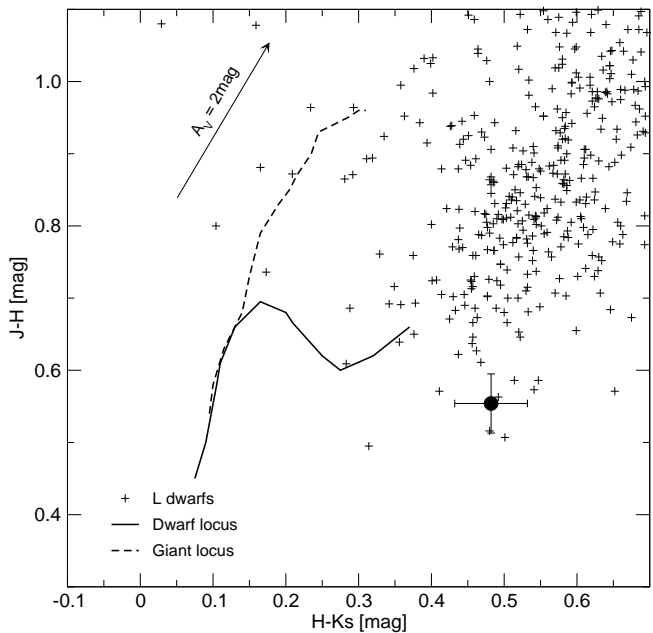}
\caption{H-Ks vs J-H comparison of the observed companion object with
  those predicted for giant and dwarf stars. A set of known L-type
  dwarfs is also shown (crosses).}
\label{jhhk}
\end{figure}

With these limitations in mind, we find that the H-Ks value observed
for this object suggests a spectral type compatible with a dwarf
star/brown dwarf, a giant star, or an extragalactic object. With the
photometric data available, the precise determination of the spectral
type, or any additional classification, is not an easy
task. Nevertheless, we can resolve whether the object is a giant star
or not. In Fig. \ref{jhhk} we see that the observed object is
compatible with a dwarf star or brown dwarf, and not a giant star in
the H-Ks vs. J-H diagram; although an extragalactic object cannot be
ruled-out.

\begin{figure}
\includegraphics[width=8.5cm]{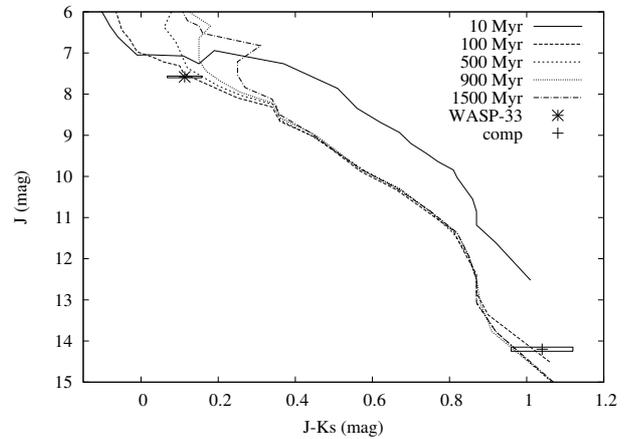}
\caption{J-Ks vs. J comparison of WASP-33 and the observed companion
  object with different isochrones.}
\label{edad}
\end{figure}

To check whether the 2 sources can be coeval, we compare their
luminosities and colours with different theoretical isochrones (Siess
et al. \cite{isoc}). Fig. \ref{edad} shows a J-Ks vs J diagram,
assuming that the 2 sources are at the same distance (116 $\pm$ 11pc;
van Leeuwen \cite{Van}). The figure shows that WASP 33 is an
early-type star with an age of 10-500 Myr, age compatible with the
spectroscopic estimation led by Cameron et al. (\cite{wasp33}). It
also shows that the NIR photometry of the candidate is that expected
for a bound low-mass star, or a substellar companion, with the same
age as WASP-33, although the age of the companion is poorer
  bounded using photometry, an age of 1500 Myr also being possible.
Contamination by a foreground or background object cannot be excluded
using photometry, except in the case of background giants that show
different NIR colours. Note that the isochrone of 100 Myr is close
to the centre of the error boxes of both objects, making this
the most probable age for both (assuming the same distance).

Confirmation of the association of WASP-33 and its tentative companion
could be made astrometrically by showing that both are common proper
motion objects. Orbital motion would be negligible, since, at a
distance of 116~pc, the projected separation of the companion is
227~AU, implying an orbital period of thousands of years.

From the revised Hipparcos catalogue, (van Leeuwen, \cite{Van}), the
absolute proper motion of WASP-33 is
($\mu_{\alpha}cos\delta=$~-1.26~mas~yr$^{-1}$,
$\mu_{\delta}=$~-9.22~mas~yr$^{-1}$), almost entirely in declination,
and the object is almost due West (positional angle 276~deg). Thus, if
the object was a distant background star with negligible absolute
proper motion then similar observations made in one year's time would
show WASP-33 having moved in declination (South) by 9.2~mas, while the
object stays still. Geometrically, the expected change in one year
with the relative position of WASP-33, and an object with a
hypothesised motion of zero, would be almost entirely in position
angle, 0.27~deg, while 0.001" in separation.

Therefore, with an uncertainty in position angle of 0.24 degrees, a
$\sim3\sigma$ detection of the relative proper motion of it and
WASP-33 could be made with a 3-year baseline, if the tentative
companion were instead a low proper motion background star. If the
tentative companion were a faint foreground star, with a consequently
large proper motion, a non-zero relative proper motion would be even
more obvious. Conversely, a zero relative position in 3 years would
indicate common proper motion and confirm that the two objects are
physical associated with one another.

\begin{figure}
\includegraphics[width=8.5cm]{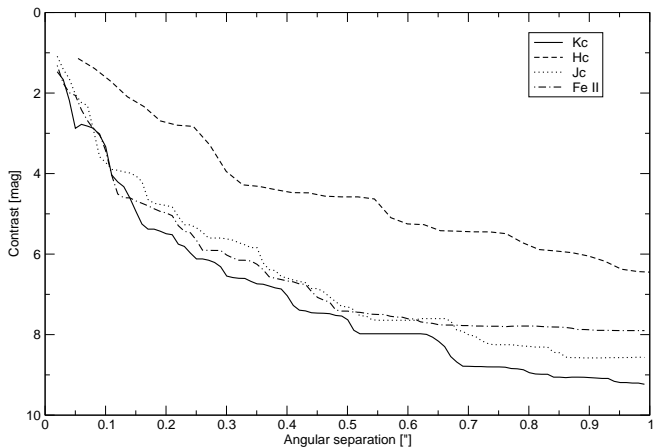}
\caption{Sensitivity of the present observations for different
    filters as a function of the angular separation. The solid line is
    the observational limit obtained with the Kc-filter,
    respectively with the dash line - Hc-filter, dot line -
    Jc-filter and, dash-dot line - Fe II filter.}
\label{sensi2}
\end{figure}

Finally, if the new discovered object is a companion of WASP-33, its
mass is in the range $[<0.1,0.2]M_\odot$ and its $T_{eff}=3050\pm250
\,^{\b{o}}K$. These ranges have been obtained comparing the observed
J-Ks vs. J photometry with the isochrones used for Fig. \ref{edad}
(Siess et al. \cite{isoc}), with ages in the range [10, 1500] Myr.

\section{Result 2: No additional objects in the image}

\begin{figure}
\includegraphics[width=8.5cm]{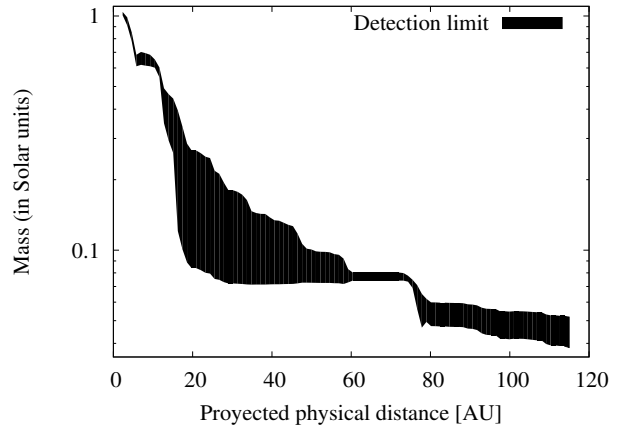}
\caption{Maximum mass in solar units of possible objects out of the
  detection limits of our observations as a function of the
    projected physical distance to the central star (in AU). The
  observational errors in the determination of the magnitude, distance
  and age have been taken into account. See text for details.}
\label{sensi1}
\end{figure}

The absence of additional objects in the image obtained offers clear
constraints to the presence of very low mass stars or brown dwarfs
orbiting WASP-33. In Fig. \ref{sensi2} we show the sensitivity of the
present observations for different filters, as a function of the
angular separation. The solid line is the observational limit obtained
with the Kc-filter, respectively the dash line - Hc-filter, dot line -
Jc-filter and, dash-dot line - Fe II filter. The Kc-filter offers the
greater accuracy and the presence of different objects can be
ruled-out as a function of the distance to WASP-33.  Fig. \ref{sensi1}
shows an estimation of this limit (using K-filter as reference) in
terms of the mass in solar units and distance from the star in AU for
the age determined for WASP-33. The isochrones used to convert
luminosities to masses are made from the latest NextGen models
(Baraffe et al. \cite{baraf1}) down to 2.300$\,^{\b{o}}$ K, and DUSTY
models (Baraffe et al. \cite{baraf2}) up to 2.300$\,^{\b{o}}$ K. The
limiting region is bounded by the errors in the determination of the
magnitude of WASP-33, its distance and its age. The age of the objects
is assumed to be in the range [10, 500] Myr, that is, from ZAMS to the
largest age estimated for WASP-33. The discontinuities in the
lower boundary of the limiting band are a reflection of the small
discontinuity found in the transition between the two models used.

In this figure, for example, we determine that objects of 0.8 $M_\odot$
can be ruled-out from existing at projected physical distances larger
than [2, 5] AU. The same can be said for 0.3 $M_\odot$ stars and
  projected physical distances larger than [11, 18] AU. Substellar
objects (masses lower than 0.072 $M_\odot$) can be ruled out at
  projected physical distances larger than 75 AU. The inner limit for
these objects cannot be properly estimated due to the discontinuity
described in the previous paragraph.

\section{Conclusions}

In the present study, high spatial resolution images of WASP-33 are
shown. We have obtained images from 1 to 2.5 $\mu$m using the
W.M. 10m-Keck II telescope and its adaptive optics system, obtaining
broad (Jc,Hc, and Kc) and narrow band (FeII) images of the system.

We have found the existence of a new object, located 1.961$"$ $\pm$
0.003$"$ distant from the central star, with a position angle 276.32
$\pm$ 0.24 deg. The colour information obtained restricts this object
to being a dwarf star/brown dwarf, or an extragalactic object. Assuming
the same distance from the Sun, we have calculated the distance
between these objects to be 227~AU. With the observations obtained, we
have also restricted the age of WASP-33 to younger than 500 Myr,
with a most probable age of 100 Myr, confirming the estimations done
in Collier-Cameron et al. (\cite{wasp33}). The companion object, under
the assumption of the same distance, presents a larger age range,
and a similar most probable age. The determination whether
  these objects are comoving or not, by using astrometry, can be
  confirmed with an additional observation, with similar accuracy,
  separated by three years.

The lack of any additional object in the image introduces constraints
on the existence of other stellar companions. With the sensitivity of
our observations, the existence of a star with a mass
0.8 $M_\odot$ at projected physical distances larger than [2, 5]
AU, 0.3 $M_\odot$ at projected physical distances larger than
[11, 18] AU and 0.072 $M_\odot$ at projected physical distances
larger than [18, 75] AU, can be ruled-out for ages in the range [10,
  500] Myr.

Our next goals are the confirmation of the link between the
discovered object and WASP-33, and the determination of its orbital
parameters. With these data we can study the possible influence of
the discovered companion on the orbit of the planet WASP-33b.

\begin{acknowledgements}
The authors want to thank the referee for useful and interesting
comments. A.M. wants to thank J.A. Caballero for fruitful
discussions and suggestions. A.M. acknowledges the funding of
AstroMadrid (CAM S2009/ESP-1496). This research has been funded by the
Spanish grants ESP2007-65475-C02-02, AYA 2010-21161-C02-02. F.M. work
was supported by the National Science Foundation under award number
AAG-0807468.
\end{acknowledgements}

\end{document}